\pdfoutput=1

\documentclass[twocolumn,showpacs,preprintnumbers,amsmath,amssymb,floatfix]{revtex4-1}

\usepackage{amsmath}
\usepackage{graphicx}
\usepackage{amssymb}
\usepackage{bm}
\begin{document}

\title{ Effect of isoscalar spin-triplet pairings on spin-isospin responses in $sd-$ shell nuclei }

\author{H. Sagawa$^{1,2)}$, T. Suzuki$^{3,4)}$, and M. Sasano$^{1)}$
}
\affiliation{
$^{1)}$RIKEN, Nishina Center, Wako, 351-0198, Japan\\
$^{2)}$Center for Mathematics and Physics, University of Aizu, Aizu-Wakamatsu, Fukushima 965-8560, Japan\\
$^{3)}$Department of Physics, College of Humanities and Science, 
Nihon Univerity, Sakurajosui 3, Setagaya-ku, Tokyo 156-8550, Japan\\
$^{4)}$National Astronomical Observatory of Japan, Mitaka, Tokyo 181-8588, 
Japan\\
%Milano and INFN, Sezione di Milano, 20133 Milano, Italy\\
%$^{4)}$Research Center for Nuclear Physics, Osaka University,
%Ibaraki, Osaka 567-0047, Japan\\
%$^{5)}$Department of Physics, Osaka University,
%Toyonaka, Osaka 560-0043, Japan\\
%%\\
%%$^{5)}$Department of Physics, Kyushu University, Higashi, Fukuoka 812-8581,
%%  Japan
}

%%%%%%%%%%%%%%%%%%%%%%%%%%%%%%%%%%%%%%%%%%%%%%%%%%%%%%%%%%%%%%
% You may repeat \author \address as often as necessary      %
%%%%%%%%%%%%%%%%%%%%%%%%%%%%%%%%%%%%%%%%%%%%%%%%%%%%%%%%%%%%%%

\begin{abstract}
The spin magnetic dipole transitions and the neutron-proton spin-spin correlations in $sd-$shell even-even nuclei with $N=Z$ are investigated using shell model wave functions.
The isoscalar spin-triplet pairing correlation provides a substantial quenching effect on the spin magnetic dipole transitions, especially the isovector (IV) ones. 
Consequently, an enhanced isoscalar spin-triplet pairing interaction influences the proton-neutron spin-spin correlation deduced from the difference between the isoscalar (IS) and the IV sum rule strengths.
The effect of the $\Delta$ ($\Delta_{33}$ resonance)-hole coupling is  examined in the IV spin transition and the spin-spin correlations of the ground states.
\end{abstract}


\maketitle
The spin-isospin response is a fundamental process in nuclear physics and astrophysics. 
The Gamow-Teller (GT) transition, which is a well-known "allowed"  charge exchange transition, involves the transfer of one unit of the total angular momentum induced by ${\vec \sigma} t_{\pm}$~\cite{Bohr1}. 
In a no-charge-exchange channel, magnetic dipole (M1) transitions are extensively observed in a broad region of the mass table.
%Both the spin and the angular momentum operators induce M1 transitions~\cite{Bohr1} as well the isovector (IV) and the isoscalar (IS) modes, depending on whether the isospin operator is included.
Both the spin and the angular momentum operators induce M1 transitions~\cite{Bohr1}, and depending on whether the isospin operator is included also induce the isovector (IV) and the isoscalar (IS) modes.
%The  M1 transitions are induced by both the spin and the angular momentum operators~\cite{Bohr1} and there are isovector (IV) and isoscalar (IS) modes, depending on whether the isospin operator is included.

Compared to the relevant theoretical predictions by shell model and random phase approximations (RPA)~\cite{Gaarde,Wakasa97,Yako05,Yako09,Sasano11,Wakasa12}, the experimental rates of these spin-isospin responses are quenched. 
%Experimental rates of these spin-isospin responses are found to be quenched compared to relevant theoretical predictions by shell model and random phase approximations (RPA)~\cite{Gaarde,Wakasa97,Yako05,Yako09,Sasano11,Wakasa12}.  
A similar quenching effect also occurs in the observed magnetic moments of almost all nuclei compared to the single-particle unit (i.e., the Schmidt value)~\cite{Bohr1, Arima,Towner}.
The quenching effect of spin-isospin excitations influences many astrophysical processes such as the mean free path of neutrinos in dense neutron matter, the dynamics and nucleosynthesis in core-collapse supernovae explosions \cite{Lan08}, and the cooling of prototype-neutron stars \cite{Reddy99}.    
Furthermore, the exhaustion of the GT sum rule is directly related to the spin susceptibility of asymmetric nuclear matter \cite{Fan2001} and the spin-response to strong magnetic fields in magnetars \cite{Radhi15}.  

Although the quenching phenomena of magnetic moments and spin responses have been extensively studied, previous research has focused mainly on the mixings of higher particle-hole (p-h) configurations \cite{Arima,Towner,BH} and the coupling to the $\Delta$ resonances \cite{BM-q,MRho}.
%The quenching phenomena in magnetic moments and spin responses have been extensively studied mainly focusing on  mixings of higher particle-hole (p-h) configurations \cite{Arima,Towner,BH} and also the coupling to the $\Delta$ resonances \cite{BM-q,MRho}.
In particular, the measured strength of the GT transitions up to the GT giant resonance is strongly quenched compared to the non-energy weighted sum rule, $3(N-Z)$~\cite{Gaarde}.
This observation has raised a serious question about standard nuclear models because the sum rule is independent of the details of the nuclear model, implying a strong coupling to $\Delta$.
After a long debate~\cite{Arima97}, experimental investigations by charge-exchange $(p,n)$ and $(n,p)$ reactions on $^{90}$Zr using multipole decomposition (MD) techniques have revealed about 90\% of the GT sum rule strength in the energy region below Ex=50~MeV~\cite{Yako05,Ichimura06}, demonstrating the significance of the $2p-2h$ configuration mixings due to the central and tensor forces~\cite{BH}, although the coupling to $\Delta$ is not completely excluded.

IV spin M1 transitions induced by ${\vec \sigma} t_z$ can be regarded as analogous to GT transitions between the same combination of the isospin multiplets.
Therefore, they should show the same quenching effect as GT transitions.
On the other hand, the IS spin M1 transitions are free from the coupling to $\Delta$ and their strength quenching should be due to higher particle-hole configurations.
Various theoretical studies have pointed out that the quenching of IS spin operators is similar to that of IV ones \cite{Richter08}.  
However, recent high-resolution proton inelastic scattering measurements at $E_p$=295~MeV have revealed that the IS quenching is substantially smaller than the IV quenching for several $N$=$Z$ $sd-$shell nuclei~\cite{Matsu2015}.

Recently, it has been reported that the isoscalar (IS) spin-triplet pairing correlations play an important role in enhancing the GT strength near the ground states of daughter nuclei with mass $N\sim Z$~\cite{Bai3,Fujita14,Tanimura14,Sagawa-Colo}.  
At the same time, the total sum rule of the GT strength is quenched by ground state correlations due to the IS pairing \cite{Zamick02}.

In this paper, we study the effect of IS spin-triplet pairing correlations on the %sum rules of 
IS and  IV spin M1 responses %as an analog transtion, and IS spin M1 responses 
 based on modern shell model effective interactions for the same set of $N$=$Z$ nuclei as those in Ref.~\cite{Matsu2015}.  
The IV response is analogous to the GT one.  
We consider that simultaneous calculations of these responses within the same nuclear model may be advantageous to distinguish the effect of the higher order configurations from  the $\Delta-$hole  coupling due to the fact that the IS spin M1 transition is independent of the $\Delta-$hole coupling strength.

We consider the IS and IV spin M1 operators, which are given as
\begin{eqnarray}
\hat{O}_{IS}&=&\sum_i{\vec \sigma}(i)  \label{IS-spin},   \\
%\hat{O}_{IV}&=&\sum_i{\vec \sigma}(i)\tau_z(i)  \label{IV-spin}, 
\hat{O}_{IV}&=&\sum_i{\vec \sigma}(i) \tau_z(i)  \label{IV-spin}, 
\end{eqnarray}
as well as the GT charge exchange excitation operators, which is expressed as
\begin{equation}
\hat{O}_{GT}=\sum_i{\vec \sigma}(i) t_{\pm}(i).
\label{ope-IV-GT}
\end{equation}
The sum rule values for the M1 spin transitions are defined by
\begin{eqnarray}
S(\vec \sigma)=\sum_f\frac{1}{2J_i+1}|\langle J_f||\hat{O}_{IS}||J_i\rangle|^2 ,\\
S(\vec \sigma \tau_z)=\sum_f\frac{1}{2J_i+1}|\langle J_f||\hat{O}_{IV}||J_i\rangle|^2 .
\end{eqnarray}
For the GT transition, the sum rule value is defined by 
\begin{eqnarray}
S(\vec \sigma t_{\pm})=\sum_f\frac{1}{2J_i+1}|\langle J_f||\hat{O}_{GT}||J_i\rangle|^2 ,
\label{GT-sumvalue}
\end{eqnarray}
and satisfies the model independent sum rule,
\begin{equation}
S(\vec \sigma t_-)-S(\vec \sigma t_+)=3(N-Z).
\label{GT-sum}
\end{equation}

According to Ref.~\cite{Matsu2015}, the proton-neutron spin-spin correlation is defined as
\begin{eqnarray}
&&\Delta_{{\rm S}}=\frac{1}{16}\left(S(\vec \sigma)-S(\vec \sigma \tau_z)\right)  \nonumber  \\
&=&
\sum_f\langle J_i|\sum_i\frac{{\vec \sigma}_n(i)+{\vec \sigma}_p(i)}{4}|J_f\rangle\langle J_f|\sum_i\frac{{\vec \sigma}_n(i)+{\vec \sigma}_p(i)}{4}|J_i\rangle       \nonumber  \\
&-& \sum_f\langle J_i|\sum_i\frac{{\vec \sigma}_n(i)- {\vec \sigma}_p(i)}{4}|J_f \rangle     
 \langle J_f|\sum_i\frac{{\vec \sigma}_n(i)-{\vec \sigma}_p(i)}{4}|J_i\rangle   \nonumber  \\
&=& \langle J_i|{\vec S_p}\cdot {\vec S_n} |J_i \rangle ,
\label{spin-spin}
\end{eqnarray}
where ${\vec S_p}=\sum_{i\in p}{\vec s_p(i)}$ and  ${\vec S_n}=\sum_{i\in n}{\vec s_n(i)}$.
The correlation value is 0.25 and $-0.75$ for a proton-neutron pair with a pure spin triplet and a singlet, respectively.  
The former corresponds to the  ferromagnet limit of the spin alignment, while the latter is the paramagnetic one.

\begin{figure}[htp]
%\vspace{-0.8cm}
\includegraphics[scale=0.3,clip,angle=-90%]{si28-is.eps}
%,clip,angle=-90
,bb=0 0 595 842]{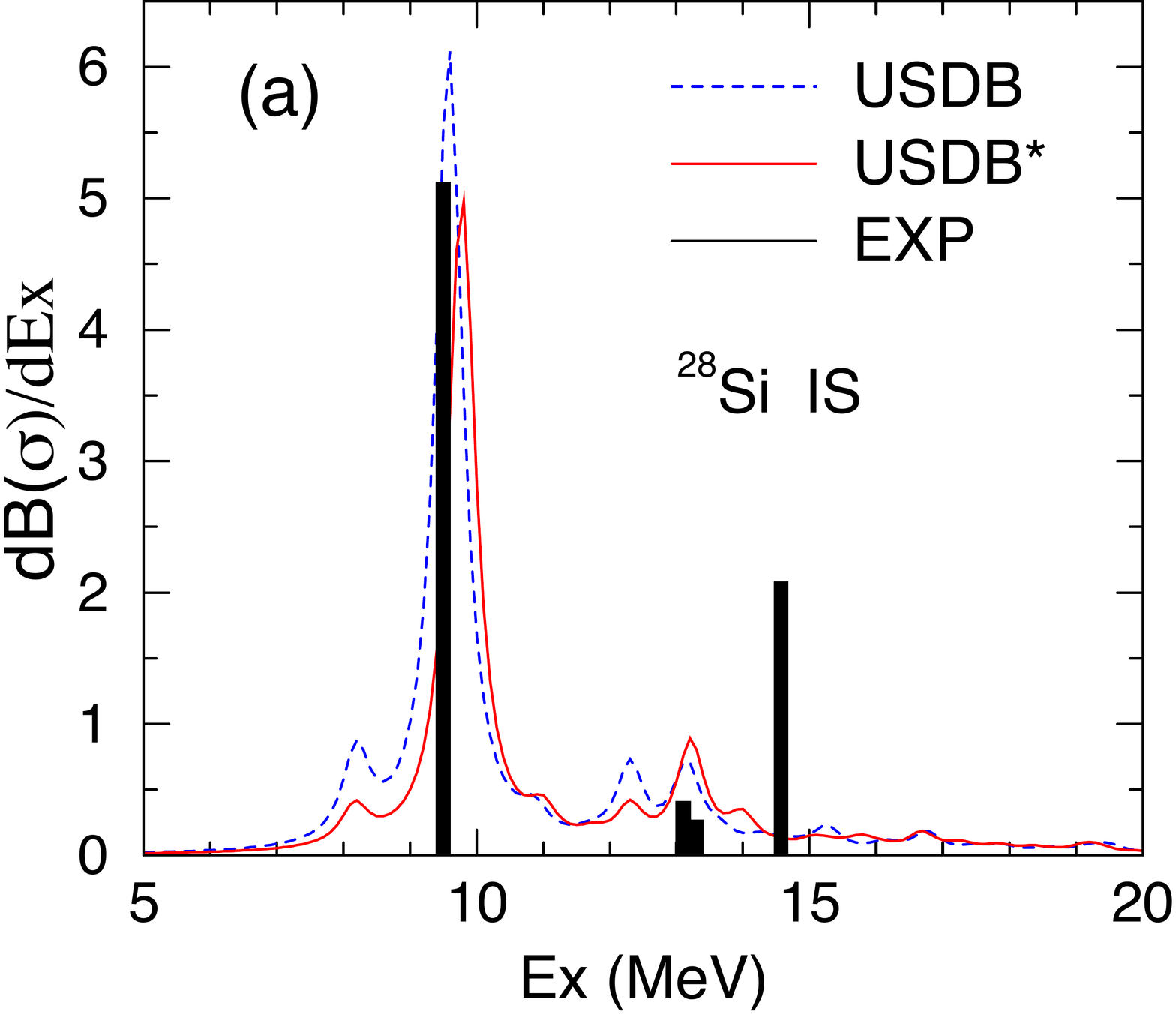}
\includegraphics[scale=0.3,clip,angle=-90%]{si28-iv.eps}
%,clip,angle=-90
,bb=0 0 595 842]{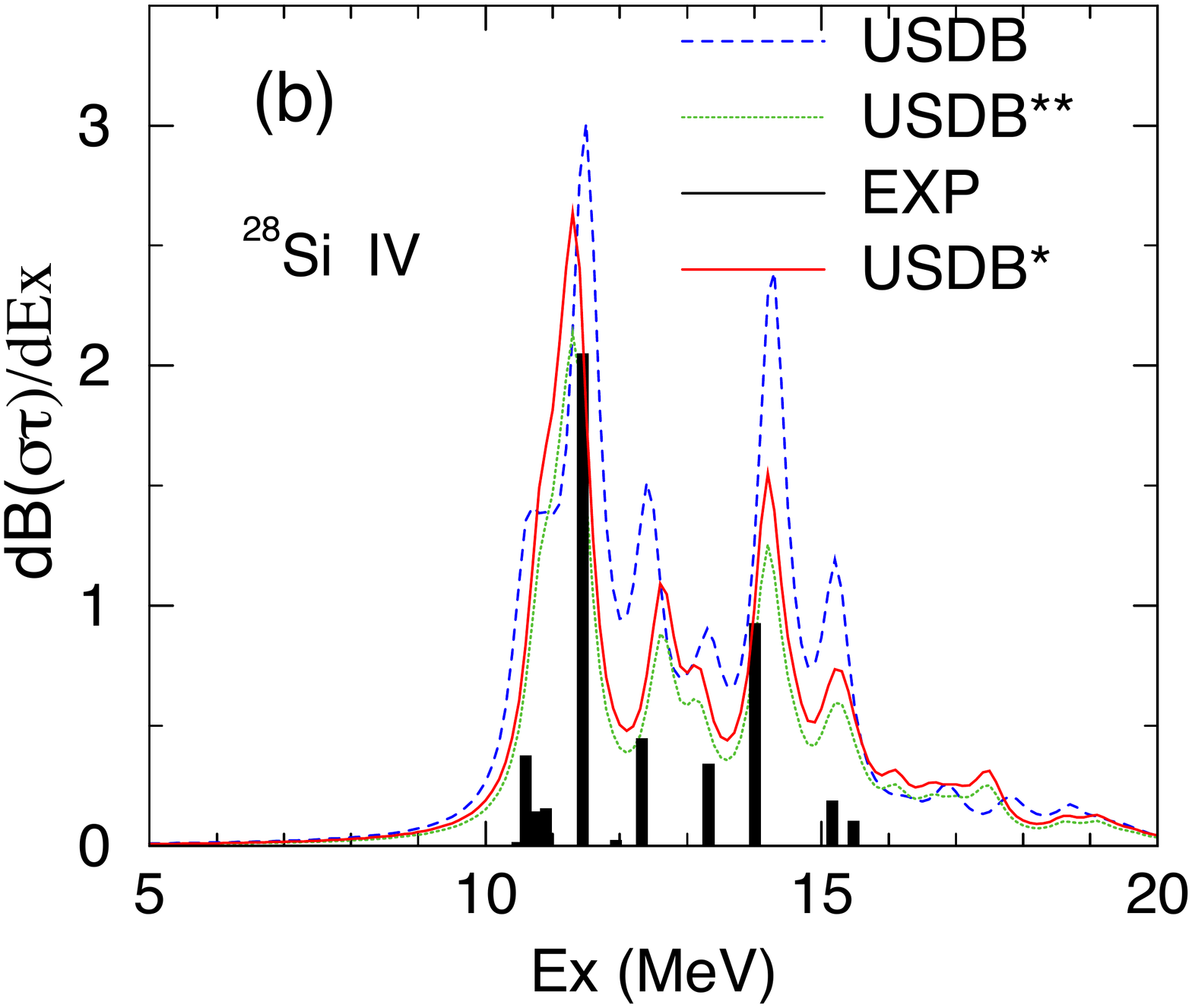}
%\vspace{-0.5cm}
\caption{(Color online) (a) IS and (b) IV spin-M1  transition strengths in $^{28}$Si.
%Experimental and theoretical data are summed up to$E_x=16$MeV .  
Shell model calculations are performed in the full $sd-$shell model space with an USDB effective interaction.  
In the results of USDB$^*$, multiplying the relevant matrix elements by a factor of 1.2 compared to the  original USDB interaction enhances the IS spin-triplet interaction.
For the results of the IV spin-M1 transitions, a quenching factor $q$=0.9 is used for USDB$^{**}$.  
Calculated results are smoothed by taking a Lorentzian weighting factor with the width of 0.5~MeV, while the experimental data are shown in the units of B($\sigma$) for the IS excitations and B($\sigma\tau$) for the IV excitations.  
Experimental data are from ref. \cite{Matsu2015}.
\label{fig:si28-m1}}
%\vspace{-0.3cm}
\end{figure}

\begin{figure}[htp]
%\vspace{-0.8cm}
\includegraphics[scale=0.3,clip,angle=-90%]{si28-is.eps}
%,clip,angle=-90
,bb=0 0 605 842]{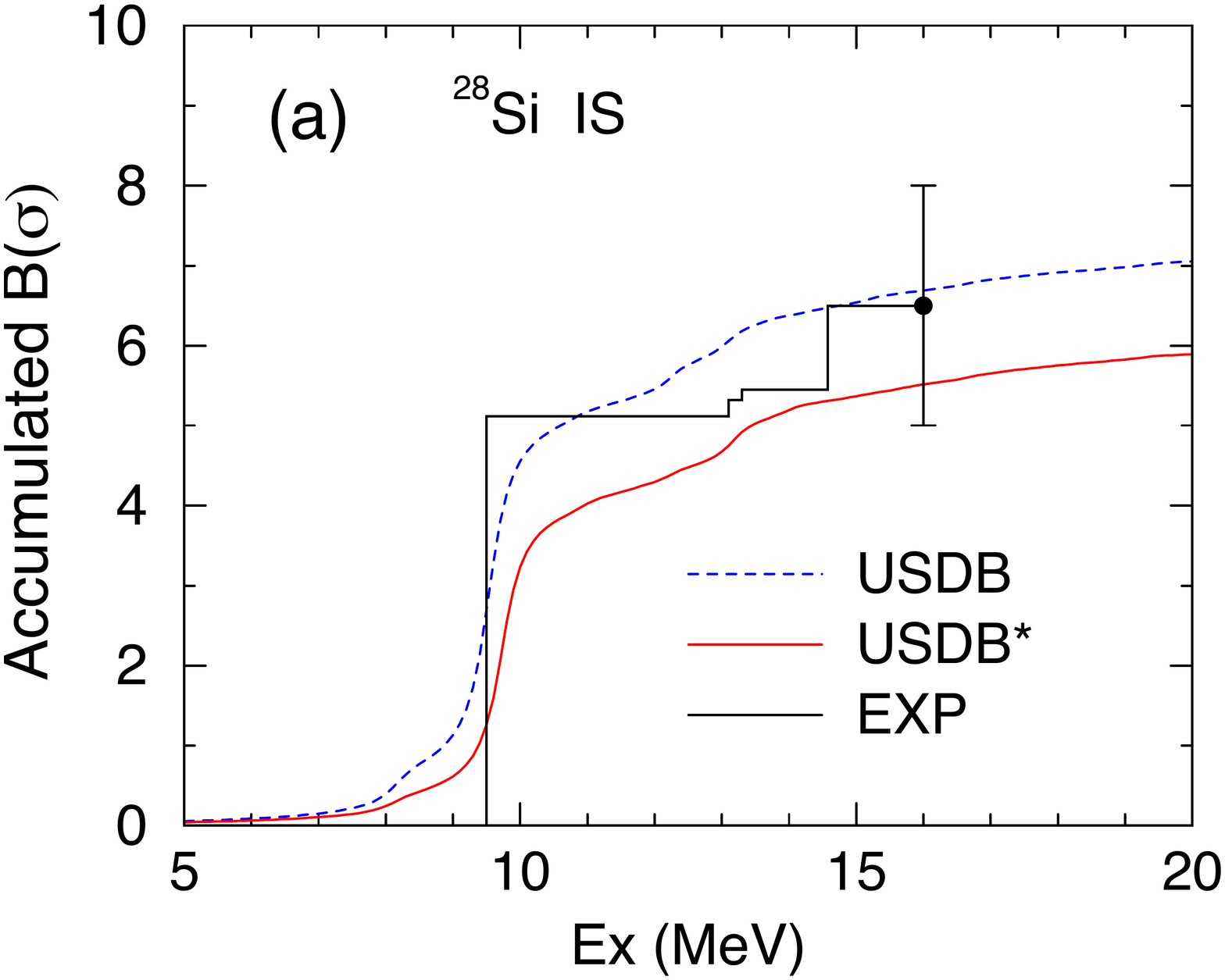}
\includegraphics[scale=0.3,clip,angle=-90%]{si28-iv.eps}
%,clip,angle=-90
,bb=0 0 595 842]{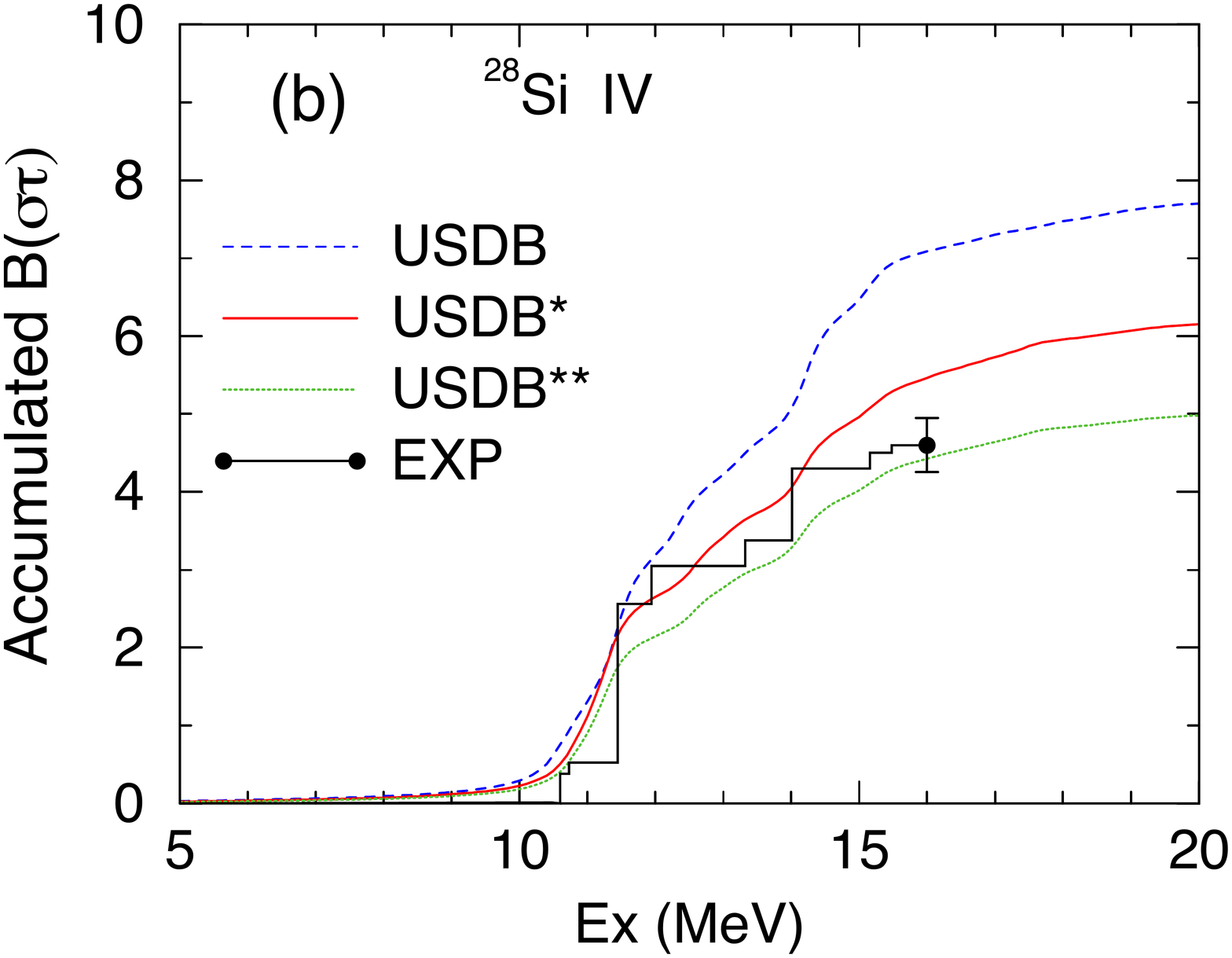}
%\vspace{-0.5cm}
\caption{(Color online) 
Accumulative sum of the IS spin-M1 strength (a) and the IV spin-M1 strength (b) as a function of the excitation energy in $^{28}$Si.
For USDB$^{**}$, a quenching factor $q$=0.9 is used for the results of the IV spin-M1 transitions.
Calculated results are smoothed in the same manner in Fig.~\ref{fig:si28-m1}.
Dot with a vertical error bar denotes the experimental accumulated sum of the strengths.
\label{fig:si28-m2}}
%\vspace{-0.3cm}
\end{figure}

The shell model calculations are performed in full $sd-$ shell model space with the USDB interaction~\cite{Brown06}. 
Among the effective interactions of the USD family, USD~\cite{Wildenthal}, USDA~\cite{Brown06}, and USDB~\cite{Brown06}, the results of spin excitations with J$^{\pi}$=1$^+$ are quite similar to each other both in excitation energies and transition strengths for collective states with large transition strengths.  
%The lowest IS 1$^+$ state in $^{20}$Ne has the largest transition strength among IS states in this nucleus and rather interaction dependent since  the transition strength is several times smaller than  strong spin transitions in other states. 
An exceptional case is the lowest IS 1$^+$ state in $^{20}$Ne.
The IS spin M1 transition from the ground state to this state, which is the largest among the transitions in this nucleus, depends on the interaction.
We attribute this behavior to the rather weak transition compared to the aforementioned collective transitions.
%has the largest transition strength among IS states in this nucleus and 

Figures \ref{fig:si28-m1} and \ref{fig:si28-m2} show the energy spectra of the %IS (IV) 
spin excitations and their accumulative sums, respectively,  in $^{28}$Si.  
The calculated results are smoothed by a Lorentzian weighting factor with the width of 0.5 MeV to guide the eye.
In the USBD$^*$ case, the IS spin-triplet matrix elements with $J^{\pi}=1^+$ and $T=0$ are enhanced by multiplying by a factor of 1.2  compared to the original USDB ones. 
Furthermore, an IV quenching factor of $q=0.9$ for the IV spin operator (\ref{IV-spin}) is introduced in the spin response of USDB$^{**}$.   
The quenching factor takes into account the $\Delta-$hole coupling effect on the IV spin transition.  

The calculations shown in the upper panels of Figs. \ref{fig:si28-m1} and \ref{fig:si28-m2} reproduce quite well the experimental IS 1$^+$ state with a strong spin transition at Ex=9.50MeV, which exhausts about 80\% of the total IS strength in  both the experiments and the calculations.  
The enhanced IS pairing has about a 20\% quenching effect on the transition strength [i.e., B($\sigma$)=6.82(5.63) in USDB(USDB$^*$)], but the excitation energies are less affected within a few hundreds keV change.  
Experimentally three other IS states are observed around Ex=(14$\sim15)$MeV  without spin assignments.   
This quenching due to the strong IS pairing corresponds to the IS spin g$_{s}$(IS) factor of g$_{s}^{eff}$(IS)/g$_{s}$(IS) =0.91, which is consistent with the quenching factor  introduced in the analysis  USDB(4) in Ref. \cite{Richter08}.

The IV spin response is shown in the lower panels of Figs.~\ref{fig:si28-m1} and  \ref{fig:si28-m2}.
Two IV 1$^+$ states with strong spin strengths of B$(\sigma\tau)$=2.05 and 0.92  are reported at  Ex=11.45 and 14.01MeV, respectively.  
The enhanced IS pairing reduces the IV spin transition strength, corresponding to the renormalization factor of g$_{s}^{eff}$(IV)/g$_{s}$(IV) =0.87.
This value is comparable to the value of 0.92 that is found as an optimal quenching parameter for magnetic moments in USDB(4)~\cite{Richter08}.
The calculated results reasonably reproduce both the excitation energies and the transition strengths in the case of USDB$^{**}$ compared to the experimental data.  
The quenching of USDB$^{**}$ in the IV spin response corresponds to g$_{s}^{eff}$(IV)/g$_{s}$(IV)=0.87$\times$0.9=0.79, which is close to the value of 0.764 found for the GT transition in USDB~\cite{Richter08}. 
Several IV states with relatively small B$(\sigma\tau)$ are also well described by the calculations.  
As a whole, the calculated strength distributions with USDB$^{*}$ and USDB$^{**}$ are more concentrated in the low energy region compared to the one with USDB. 
This behavior may be considered as the same effect studied in the $fp$ shell region using the RPA  framework with the IS  pairing effect in the final state~\cite{Bai3}.
The energy spectra of other N=Z even-even nuclei (i.e., $^{24}$Mg, $^{32}$S and $^{36}$Ar) are also reproduced  quite well; 
both the excitation energies as well as the transition strengths have the same quantitative level as those of $^{28}$Si \cite{SS06}.

\begin{figure}[t]
%\vspace{-0.8cm}
\includegraphics[scale=0.3,clip,angle=-90%]{spin-is.eps}
%,clip,angle=-90
,bb=0 0 595 842]{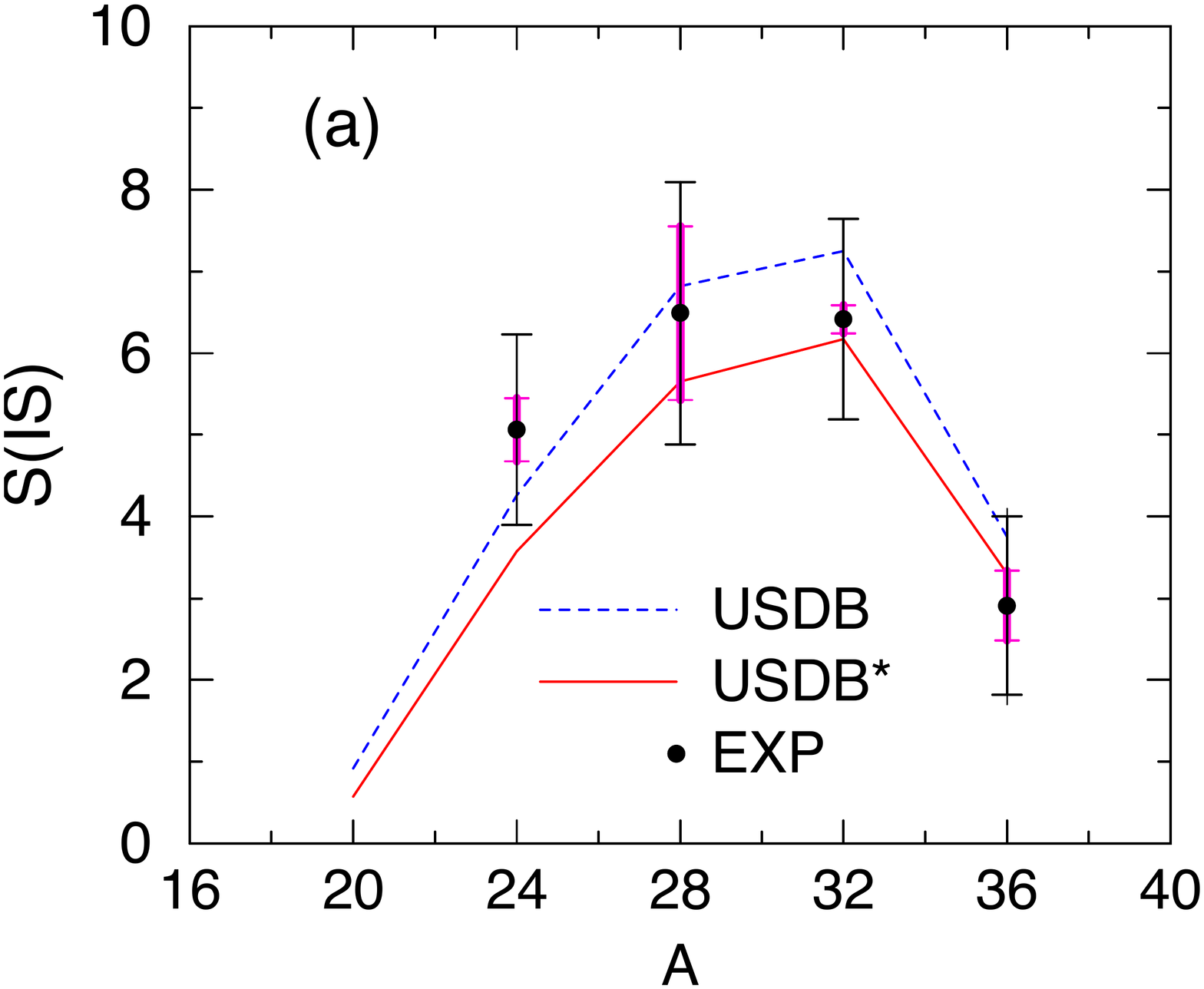}
\includegraphics[scale=0.3,clip,angle=-90%]{spin-iv.eps}
%,clip,angle=-90
,bb=0 0 595 842]{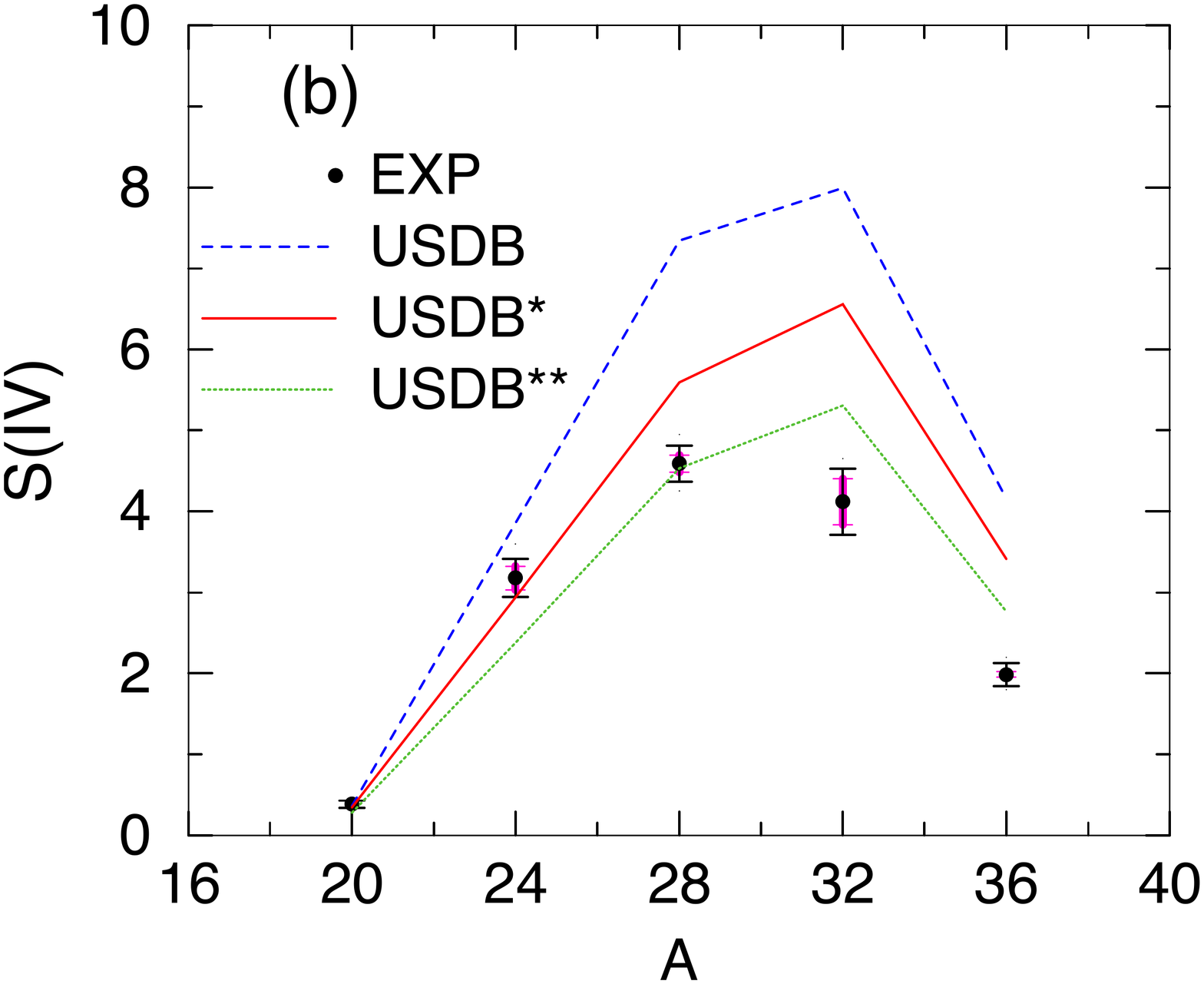}
\vspace{-0.5cm}
\caption{(Color online) Sums of the spin-M1 transition strengths of IS (a) and IV (b).
Experimental and theoretical data are summed up to Ex=16MeV .  
Shell model calculations are performed with the USDB effective interaction.  
In the results of USDB$^*$, the IS spin-triplet interaction is enhanced by multiplying the relevant matrix elements by a factor of 1.2 compared to the  original USDB interactions.  
For USDB$^{**}$, a quenching factor $q$=0.9 is used for the results of the IV spin-M1 transitions.
Experimental data are from ref. \cite{Matsu2015}.
Long thin error bars indicate the total experimental uncertainty, while the short thick error bars denote the partial uncertainty from the spin assignment.  \label{fig1}}
%\vspace{-0.3cm}
\end{figure}

Figures~\ref{fig1} (a) and (b) show the sum rule values of $S(\vec \sigma)$ and $S(\vec \sigma \tau_z)$  for the USDB and USDB$^*$ interactions, respectively.   
A strong IS spin-triplet correlation in the ground states suppresses the IV sum rule more than the IS one in the comparison between USDB and USDB$^*$.
On top of the strong IS pairing, a quenching factor $q=0.9$ on the IV spin operator (\ref{IS-spin}) is used for USDB$^{**}$  to simulate the coupling to the $\Delta$ state, which may affect only the isovector sum rule due to the isovector nature of the $\Delta-$hole excitations.

\begin{figure}[t]
%\vspace{0.5cm}
\includegraphics[scale=0.3,clip,angle=-90%]{delta-s.eps}
%,clip,angle=-90
,bb=0 0 595 842]{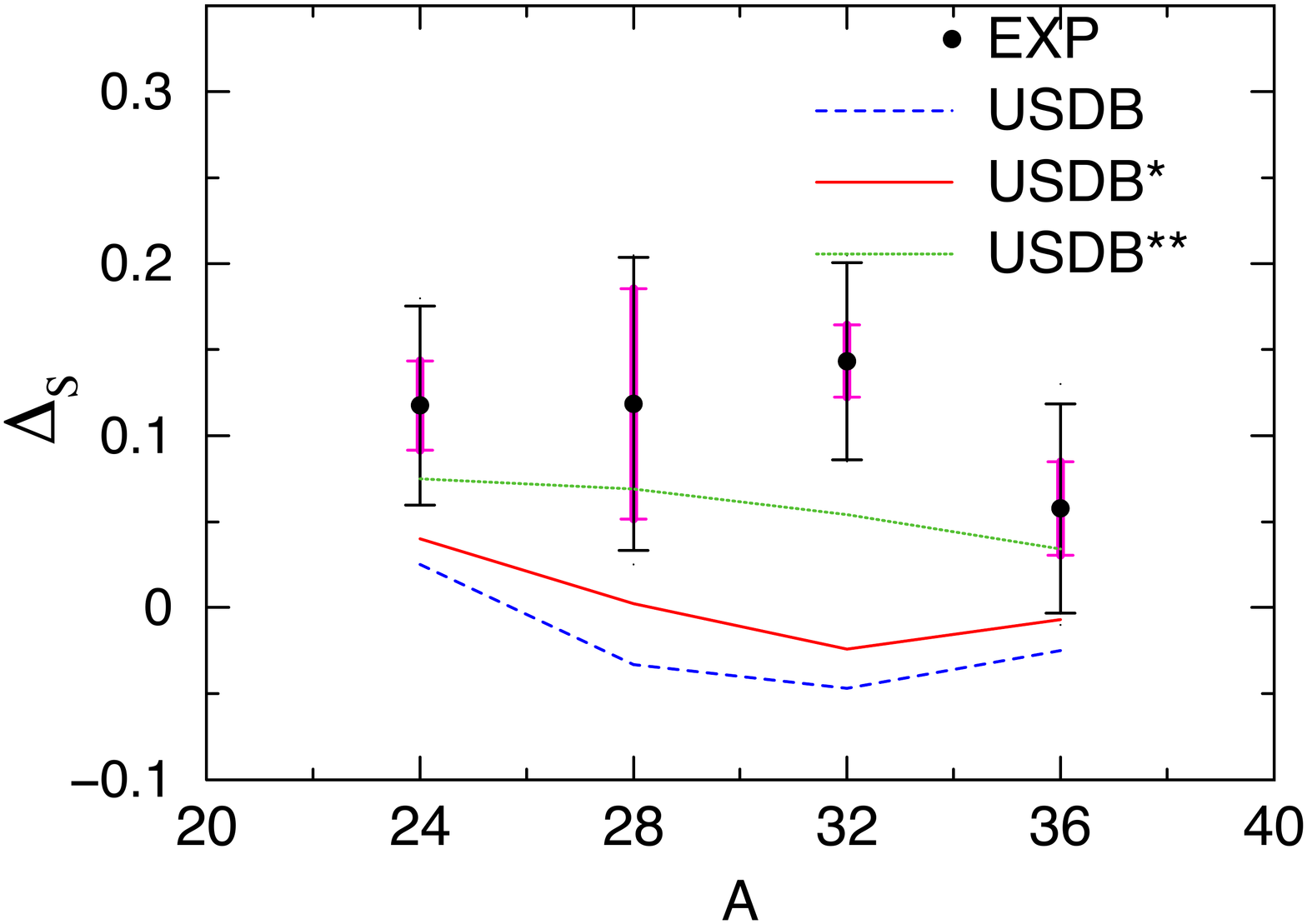}
%\vspace{-0.5cm}
\caption{(Color online) 
Experimental and calculated proton-neutron spin-spin correlation $\Delta_{{\rm S}}$.  
Spin M1 transition strengths are summed up to Ex=16MeV.  
Shell model calculations are performed with an effective interaction USDB.  
In the results of USDB$^*$, the IS spin-triplet interaction is enhanced by multiplying the relevant matrix elements by a factor of 1.2 compared to the original USDB.  
A quenching factor  $q$=0.9 is used for the results of the IV spin-M1 for  USDB$^{**}$.  
Experimental data are taken from ref. \cite{Matsu2015}. 
See the caption of Fig. \ref{fig1} for a description of the experimental error bars. 
\label{fig:spin-spin}}
%\vspace{-0.3cm}
\end{figure}

Figure \ref{fig:spin-spin} shows the experimental and the calculated proton-neutron spin-spin correlations (\ref{spin-spin}).
Although the experimental data still have large error bars, the calculated results with the USDB interaction show poor agreement with the experimental data.  
This is also the case for the other USD interactions such as USD and USDA.  
The results with an enhanced IS spin-triplet pairing improve the agreement appreciably.
As a result, a quenching factor close to unity, $q=0.9$, eventually results in a fine agreement with the experimental observations.
The positive value of the correlation indicates that the population of spin triplet pairs in the ground state is larger than that of the spin singlet pairs.

To clarify the physical mechanism of the IS spin-triplet interaction, we make a perturbative treatment of the 2-particle 2-hole (2p-2h) ground state correlations on the spin-spin matrix element.   
We express the wave function for the ground state with proton-neutron correlations for even-even N=Z nuclei, $|\tilde{0}\rangle$, as 
%\begin{widetext}
\begin{eqnarray}
&&|\tilde{0}\rangle=|0\rangle 
+\sum_{i=1,2,1',2'}\alpha(1,2,1',2')  \nonumber \\
&\times& |(1_{\pi}2_{\nu})J_1,T_1;(1^{'-1}_{\pi}2_{\nu}^{'-1})J_2,T_2:J=T=0\rangle.
\label{eq:pt1}
\end{eqnarray}
%\end{widetext}
Here the first term on the right hand side, $|0\rangle$, is the wave function with no finite seniority $\nu=0$ (i.e., without the spin-triplet correlations).
The second term represents the states of 2-particle $(1_{\pi}2_{\nu})$ and 2-hole $(1^{'-1}_{\pi}2_{\nu}^{'-1})$ for a proton ($1_{\pi}$ or $1^{'}_{\pi}$) and neutron $(2_{\nu}$ or $2^{'}_{\nu})$ pair.
The indices $(i\equiv 1,2,1',2')$ stand for the quantum numbers of the single-particle state $i=(n_i,l_i,j_i)$.  
In Eq. (\ref{eq:pt1}), the perturbative coefficient is given by 
\begin{eqnarray}
&&\alpha(1,2,1',2')   \nonumber \\
&=&\frac{\langle (1_{\pi}2_{\nu})J_1,T_1;(1^{'-1}_{\nu}2_{\pi} ^{'-1})J_2,T_2:J=T=0|
H_{p}|0\rangle}{\Delta E}   \nonumber \\
\end{eqnarray}
where $H_p$ is the IS spin-triplet two-body pairing interaction and $\Delta E =E_0-E(12;1^{'-1}2^{'-1})$.
%The 2-particle 2-hole states are seniority $\nu$=4 states in Eq. (\ref{eq:pt1}).  
The 2p-2h states are the seniority $\nu$=4 states in Eq. (\ref{eq:pt1}).  
Since the pairing interaction $H_p$ is attractive and the energy denominator $\Delta E$ is negative,  the perturbative coefficient $\alpha(1,2,1',2')$ shoud be positive.  
The effect of the ground state  correlations on the proton-neutron spin-spin matrix is then  evaluated as
%\begin{widetext}
\begin{eqnarray}
&&\langle \tilde{0}|{\vec S_p}\cdot {\vec S_n} | \tilde{0}\rangle 
= 2\sum_{1,2,1',2'}\alpha(1,2,1',2') \nonumber \\ %\langle0|{\vec S_p}\cdot {\vec S_n}|(1_{\pi}2_{\nu})J_1=1T_1=0;(1^{'-1}_{\nu}2_{\pi} ^{'-1})J_2=1,T_2=0:J=T=0\rangle
&\times&\langle0|{\vec S_p}\cdot {\vec S_n}|(1_{\pi}2_{\nu})J_1,T_1;(1^{'-1}_{\nu}2_{\pi} ^{'-1})J_2,T_2:J=T=0\rangle  \nonumber \\
\label{eq:pt2}
\end{eqnarray}
%\end{widetext}
where the angular momenta and the isospins are selected to be $J_1=J_2=1$ and $T_1=T_2=0$  by the nature of %the neutron-proton spin-spin matrix.
%That is, in the perturbative formula (\ref{eq:pt2}), 
the IS spin-triplet interaction.
The matrix element in Eq. (\ref{eq:pt2}) is further expressed as a reduced matrix element in the spin space,
%\begin{widetext}
\begin{eqnarray}
&&\langle 0||{\vec S_p}\cdot {\vec S_n}||(1_{\pi}2_{\nu})J_1;(1^{'-1}_{\nu}2_{\pi} ^{'-1})J_2:J=0\rangle   \nonumber \\
&=&\delta_{J_1,J_2} \delta_{J_1,1}\sum_{J'}(-)^{j_1+j_2+J_1+J'}\left\{\begin{array}{rrr}
j_1 &j_2&J_1  \\
j_2' &j_1' &J' 
\end{array} \right\}  \nonumber \\
&\times&\langle 0||{\vec S_p}\cdot {\vec S_n}||(1_{\pi}1_{\pi}^{'-1})J';(2_{\nu}2_{\nu} ^{'-1})J':J=0\rangle  
\nonumber \\
&=& \frac{1}{\sqrt{3} }  \left\{\begin{array}{rrr}
j_1 &j_2&J_1  \\
j_2' &j_1' &J'
\end{array} \right\}  
(-)^{j_2+j_2'}\langle j_1'||{\vec s_p}||j_1\rangle 
  \langle j_2'||{\vec s_n}||j_2\rangle   \nonumber \\
\label{eq:redmat}
\end{eqnarray}
%\end{widetext}
where the $6j$ symbol is used to evaluate the reduced matrix element.   
In  Eq.(\ref{eq:redmat}), the coupled angular momentum $J'$ is taken as $J'=1$ due to the selection rule of the spin matrix element.  
The isospin quantum number is discarded since it gives a trivial constant in Eq. (\ref{eq:redmat}).
%The spin-spin matrix element is evaluated by the formula for the spin matrix 
%\begin{eqnarray}
%&&\langle j'||{\vec \sigma}||j\rangle =\delta_{l,l'}  \nonumber \\
% &\times&\begin{cases}
%   \sqrt{\frac{2j+1}{j(j+1)}}  \times \begin{cases}   j+1 &  {\rm for} \,\,j=j', \,\, j_>=l+1/2  \\
 %                                          -j    & {\rm for}\,\, j=j', \,\,j_<=l-1/2     \end{cases}    & %\rm{for}  \,\,\,j=j'  
 %   \\
%     \mp2\sqrt{\frac{2l(l+1)}{2l+1}}    \hspace{1cm}\,\,\,\,{\rm for} \,\,\,  j'=j\pm 1  &  \end{cases}  %\nonumber
%\label{eq:redmat1}
%\end{eqnarray}
%where we use the $l\cdot s$ coupling scheme for the single-particle wave function.  
We can obtain the effect of the IS spin-triplet pairing correlations  on  the proton-neutron spin-spin correlation matrix element using the one-body spin matrix element %(\ref{eq:redmat1}) 
and the $6j$ symbol.  
It is shown that relevant  matrix elements (\ref{eq:redmat}) for 2p-2h configurations with $j_1=j_2$ and $j_1'=j_2'$  are  positive values for different combinations of $(j_1, j_1')$ [i.e,  $(j_1=j_1'=j_>=l+1/2)$,   $(j_1= j_1'=j_<=l-1/2)$,  $(j_1=j_>, j_1'=j_<)$ or $(j_1=j_<, j_1'=j_>)$].  
Thus, the numerical results shown in Figs. 1 and 2 can be qualitatively understood by using these formulae for 2p-2h configuration mixing due to the IS spin-triplet pairing.  

%% The quenching of the total GT strength from the sum rule value $S(\sigma t_-)-S(\sigma t_+)=3(N-Z)$ in Eq. (\ref{GT-sum}) has been studied theoretically due to the couplings to higher 2p-2h configurations and also $\Delta -$h excitations.  
%% Theoretical studies based on a quark model suggested about (30-40)\% quenching effect on the M1 and GT transition strength due to the $\Delta -$h couplings to the GT p-h giant resonances \cite{Bohr1}.   
%% However, experimental investigations by $(p,n)$ and $(n,p)$ reactions on $^{90}$Zr using the multipole decomposition (MD) techniques reveal about 90\% of the GT sum rule strength in the energy region below E$_{\rm x}$=50MeV.  
%% That is, 
As mentioned in the introduction, 2p-2h configuration mixings are the dominant effect for the quenching of the GT giant resonance peak, while $\Delta -$h coupling plays a minor role at most  10\%  of the effect on the sum rule value in $^{90}$Zr. 
MD analysis is also performed for $(p,n)$ reactions on $^{208}$Pb, and a quenching factor $q^2=0.86$ provides a quantitative agreement between the RPA calculations and the observed GT strength for the GT giant resonance \cite{Wakasa12}. 
The present analysis of the IV spin M1 strength suggests that  a quenching factor $q^2=(0.9)^2=0.81$ for the IV transition is necessary  to realize quantitative agreement  with the observed  spin strength of N=Z even-even nuclei.
%This quenching might be considered mainly due to the $\Delta-h$ coupling to the GT states because of IV nature of the effect although other effects such as multi-particle muclti-hole excitations are not exculsive.  
This quenching effect of 20\% gives the upper limit of the effect of the $\Delta-h$ coupling to the GT states because other effects such as multi-particle multi-hole excitations are not exclusive.
This upper limit is considerably smaller than the suggested value of 30$\sim$40\%, by the quark model~\cite{BM-q}.   
%Theoretical studies based on a quark model suggested about (30-40)\% quenching effect on the M1 and GT transition strength due to the $\Delta -$h couplings to the GT p-h giant resonances \cite{Bohr1}.   
To validate this lower quenching effect, more comprehensive and less model dependent methods should be experimentally and theoretically studied in the future.

The IS and IV spin quenching factors are traditionally evaluated by using IS magnetic moments and $\beta$ decay rates \cite{BW83}.  We check the difference between USDB and USDB* for these observables with $T=1/2$ sd shell nuclei.  We found that both interactions give quite reasonable results in comparisons with available experimental data and the differences are  quite small for both  observables. For $\beta$ decay rates  in the nuclei  with $T=1/2$, the initial and the final states involve an unpaired nucleon which  masks the pairing effect, while all the nucleons are coupled with either $J=0$ or $J=1$ in the main configurations of both the initial and the final states in the present study.  These paired configurations in even-even $N\sim Z$ mother states get the  maximum IS pairing effect \cite{Bai3,Sagawa-Colo} so that the IS and IV M1 sum rule values are substantally quenched by a stronger IS pairing  in USDB*. The same quenching effect is found on the sum rule values of GT transitions from a mother nucleus $^{26}$Mg to a daughter nucleus $^{26}$Al, especially , in the transitions to the final states with isospin $T_f$=1, and 2.  

The meson exchange currents (MEC) and  configuration mixings higher than $2\hbar \omega$ might also contribute to the renormalizations of spin and spin-isospin operators.
The MEC effect is small for IS M1 and GT transitions, while it is about 10\% 
effect on IV M1 transition matrix in $sd-$shell nuclei.  
The higher cofiguration mixing effect has an opposite sign to the MEC effect on IV M1 and tends to cancel each other \cite{BW83}.  These effects should be examined in details in future study 
together with the IS pairing effect.

In summary, we studied the IS and IV spin M1 transitions in even-even N=Z $sd-$shell nuclei using shell model calculations with  USDB interactions in full $sd-$shell model space.    
In general, the calculated results show a reasonable agreement with the experimental energy spectra with respect to both the excitation energy and the transition strengths.  
The quenching of the spin M1 transitions is obtained by  an enhanced  IS spin-triplet pairing correlation instead of using effective operators with quenched g$_{s}$ factors on top of the original USDB interaction,  without significantly changing the excitation energies themselves. 
In particular, the  quenching effects on the spin M1 transition matrices are larger on the IV spin ones  than the IS ones.  
Positive contributions for the spin-spin correlations are also found by an enhanced isoscalar spin-triplet pairing interaction in these $sd-$shell nuclei.  
The effect of the $\Delta-$hole coupling is also examined on the IV spin transition, and the empirical spin-spin correlations in the ground states are reproduced well by a combined effect of the IS pairing and a quenching factor of $q$=0.9 on the IV spin transition matrix elements.

%\begin{acknowledgments}
%\vspace{-0.3cm}
We would like to thank  H. Matsubara for providing the experimental data.
We would also like to recognize A. Tamii,  H. Sakai, and T. Uesaka for the useful discussions.
This work was supported in part by JSPS KAKENHI  Grant Numbers JP16K05367 and JP15K05090.
% from the MEXT of Japan.
%under the grant nunmber  *****.
%\end{acknowledgments}


\vspace{-0.3cm}

\begin{thebibliography}{99}
\bibitem{Bohr1} A. Bohr and B. R. Mottelson, {\it Nuclear Structure} (World Scientific, 1969), Vol. I.
%\bibitem{Bohr1} A. Bohr, B. R. Mottelson, D. Pines, Phys. Rev. 110, 936 (1958).
\bibitem{Gaarde} C. Gaarde et al.,   Nucl. Phys. A369 258 (1981). 
\bibitem{Wakasa97} T. Wakasa et al., Phys. Rev. C 55, 2909 (1997).
\bibitem{Yako05} K. Yako et al., Phys. Lett. B615, 193(2005). 
\bibitem{Yako09} K. Yako et al.,  Phys. Rev. Lett. 103, 012503 (2009).
\bibitem{Sasano11} M. Sasano, {\it et al.}, Phys. Rev. Lett. 107, 202501 (2011);\\
M. Sasano et al., Phys. Rev.  C86, 034324 (2012).  
\bibitem{Wakasa12} T. Wakasa et al., Phys. Rev. C 85, 064606 (2012).
\bibitem{Arima}  A. Arima , K. Shimizu, W. Bentz and H. Hyuga,  Adv. Nucl. Phys. 18, 1 (1987). 
\bibitem{Towner} I. S. Towner and F. C. Khanna, Nucl. Phys. A 399, 334 (1983);  
 I. S. Towner,  Phys. Rep. 155, 263 (1987).
\bibitem{Lan08}  K. Langanke et al., Phys. Rev. Lett.  93, 202501 (2004); 100, 011101 (2008).
\bibitem{Reddy99}  S. Reddy  et al., Phys. Rev. C59, 2888(1999); A. Burrows and R. F. Sawyer, Phys. Rev. C58, 554 (1998).  
\bibitem{Fan2001}  S. Fantoni et al., Phys. Rev. Lett. 87, 181101 (2001); G. Shen et al., Phys. Rev. C87, 025802 (2013).
\bibitem{Radhi15}  A. Radhi et al., Phys. Rev. C91, 045803(2015).  
\bibitem{BH}   G. Bertsch and I. Hamamoto,  Phys. Rev. C26, 1323(1982).  
\bibitem{BM-q}  A. Bohr and B. R. Mottelson,  Phys. Lett. 100B, 10 (1981).
\bibitem{MRho}  M. Rho, Nucl. Phys. A 231, 493 (1974); E. Oset and M. Rho, Phys. Rev. Lett. 42, 47 (1979).
\bibitem{Arima97} A. Arima, in {\it Proc. of the Int. Symp. on New Facet of Spin Giant Resonances in Nuclei, 1997}, (edited by H. Sakai, H. Okamura, and T. Wakasa) % (Univ. of Tokyo, Japan, 1997),
 p. 3.
\bibitem{Ichimura06}  M.  Ichimura, H. Sakai and Wakasa,    Prog. Part. Nucl. Phys.  56, 446 (2006).  
\bibitem{Richter08}  W. A. Richter, S. Mkhize and B. A. Brown,  Phys. Rev C78, 064302 (2008).
\bibitem{Matsu2015}   H. Matsubara et al., Phys. Rev. Lett.115, 102501 (2015) and private communications.  
\bibitem{Bai3} C.L. Bai, H. Sagawa, M. Sasano, T. Uesaka, K. Hagino, 
H.Q. Zhang, X.Z. Zhang, and F.R. Xu, Phys. Lett. B719, 116 (2013). 
\bibitem{Sagawa-Colo} H. Sagawa, C.L. Bai and G. Col\`o,  Physica Scripta 91, 083011  (2016) %the '75 Nobel Prize celebration volume(2016) in press.  
%\bibitem{Bohr2} A. Bohr, B. R. Mottelson, Nuclear Structure, Vol. I, Benjamin, New York, 1969.
\bibitem{Fujita14} Y. Fujita {\it et al.}, Phys. Rev. Lett. 112, 112502 
(2014).
\bibitem{Tanimura14} Y. Tanimura, H. Sagawa and K. Hagino, 
Prog. Theor. Exp. Phys. 053D02 (2014).
\bibitem{Zamick02} S. J. Q. Robinson and L. Zamick,  Phys. Rev. C66, 034303 (2002).
\bibitem{Brown06} B. A. Brown and W. A. Richter, Phys. Rev. C74, 034315 (2006).
\bibitem{Wildenthal} 
B. H. Wildenthal, Prog. Part. Nucl. Phys. 11, 5 (1984); 
B. A. Brown and B. H. Wildenthal, Annu. Rev. Nucl. Part. Sci. 38, 29 (1988).
\bibitem{SS06}   H. Sagawa and T. Suzuki, to be published.
%\bibitem{Wakasa12} T. Wakasa {\it et al.}, Phys. Rev. C85, 064606 (2012).
\bibitem{BW83}  B. A. Brown and B. H. Wildenthal, Phys. Rev. C28, 2397 (1983).  

\end{thebibliography}
\end{document}